\documentstyle[12pt]{article}
\def \sign{{\rm sign}}

\newcommand{\n}{\frac{1}{n}}
\newcommand{\nn}{\frac{1}{n-1}}
\newcommand{\noi}{\noindent}
\newcommand{\eq}{\begin{equation}}
\newcommand{\en}{\end{equation}}
\newcommand{\eqa}{\begin{eqnarray}}
\newcommand{\ena}{\end{eqnarray}}

\newcommand{\aleq}{\mbox{}_{\textstyle \sim}^{\textstyle < }}
\newcommand{\ageq}{\mbox{}_{\textstyle \sim}^{\textstyle > }}
\newcommand{\ra}{\rightarrow}
\newcommand{\be}{\begin{equation}}
\newcommand{\ee}{\end{equation}}
\newcommand{\bea}{\begin{eqnarray}}
\newcommand{\eea}{\end{eqnarray}}
\begin{document}
\hbox{}
\noindent September 1992 \hfill HU Berlin-IEP-92/5
\begin{center}
\vspace*{1.5cm}

\renewcommand{\thefootnote}{\fnsymbol{footnote}}
\setcounter{footnote}{1}

{\LARGE \bf Modified $U(1)$ lattice gauge theory:
towards realistic lattice QED}
\footnote{Work supported by the Deutsche
Forschungsgemeinschaft under research grant Mu 932/1-1 } \\

\vspace*{1.5cm}
{\large
V.G.~Bornyakov $\mbox{}^1$,
V.K.~Mitrjushkin $\mbox{}^2$ \footnote{Permanent adress:
Joint Institute for Nuclear Research, Dubna, Russia},
M.~M\"uller-Preussker $\mbox{}^2$,
}\\
\vspace*{0.7cm}
{\normalsize
$\mbox{}^1$ {\em Institute for High Energy Physics                        \\
                 142284 Protvino (Moscow Region), Russia                 }\\
$\mbox{}^2$ {\em Humboldt-Universit\"{a}t zu Berlin, Fachbereich Physik   \\
                 Institut f\"ur Elementarteilchenphysik                   \\
                 Invalidenstr. 110, O-1040 Berlin, Germany              }}\\
\vspace*{2cm}
{\bf Abstract}
\end{center}
We study properties of the compact $~4D~$ $U(1)$ lattice
gauge theory with monopoles {\it removed}.
Employing Monte Carlo simulations we calculate correlators
of scalar, vector and tensor operators at zero and nonzero
momenta $~\vec{p}~$.
We confirm that the theory without monopoles has no phase
transition, at least, in the interval $~0 < \beta \leq 2~$.
There the  photon becomes massless
and fits the lattice free field theory
dispersion relation very well.
The energies of the  $~0^{++}~$, $~1^{+-}~$ and $~2^{++}~$ states
show a rather weak dependence on the coupling in the interval
of $~\beta~$ investigated,
and their ratios are practically constant.
We show also a further modification of the theory
suppressing the negative plaquettes
to improve drastically the overlap with the lowest states
(at least, for $~J=1$).

\newpage
\section{Introduction}

Despite the big efforts invested last years into the
nonperturbative study of QED (for reviews see, for instance,
\cite{qed1,qed2})
we still have no clear understanding of this theory.
For the standard compact formulation there is a strong
indication for a
first order phase transition, thus leaving no room for the continuum
limit. The noncompact formulation seemed to be more promising
because the (chiral) phase transition turned out to be
of second order.
However, the noncompactness is realized only in the pure
gauge part of the action, while in the fermionic part
a compact $~U(1)~$ phase remains.
This part of the action will
eventually induce the compact term back into the action,
thus living us with the question of selfconsistency of the (so-called)
noncompact theory. Moreover, the compact part of the action
is responsible for the appearance of monopoles, which are
lattice artifacts and are supposed to be at least in part
the driving force of the chiral phase transition.
If the latter is correct then the chiral phase transition
in the noncompact QED might not have a continuum analogue
\cite{wang}.

The problem of lattice artifacts should be taken very
seriously in lattice calculations.
Within the Monte Carlo approach, in order to calculate
the average of any field
operator one generates a sequence of (equilibrium) gauge field
configurations:  $\{ U_{link}^{(1)} \}$,
$\{ U_{link}^{(2)} \}, \quad \ldots ~$
with the statistical weight
$P(U_{l}) \sim \exp \left[- S(U_l) \right]~$,
where $~S(U_{l})~$ is the Euclidean lattice discretized action,
and averages over all of them.
The crucial question in this approach is, whether
in a certain range of coupling(s)
these field configurations can provide an adequate representation
of continuum physics.
It is a well-known fact that artifacts can change the structure
of a theory drastically.
One example of the influence of lattice artifacts
is provided by the study of topological properties of
gauge theories. Small--scale fluctuations, i.e., the fluctuations
living on the
scale size of, say, one lattice spacing, and carrying a non-trivial
topological charge (dislocations) can lead to the divergence
of the topological susceptibility in the continuum limit (at least
for the geometric definition of the topological charge).
A special source of troubles are field configurations
containing negative plaquettes: $~\frac{1}{2}{\rm Tr} U_{P} \simeq -1~$
(for $~SU(2)~$) \cite{pt1,pt2,desy1}.
A similar situation takes place for the $~CP^{n-1}~$ model with
$~n \leq 3~$ \cite{lus2}.
 The fact that singular configurations may be important in
quantum field theory was demonstrated also for the
three--dimensional non--linear $~O(3)~$ model \cite{hkp}
and for the compact Sine-Gordon model \cite{hp} .
It was speculated that point--like singularities (topological
defects) are the sources of a topological anomaly, and a new
coupling constant controls their behaviour.
This `hidden' coupling constant without
making important contributions to the short--distance behaviour
becomes crucial in understanding the infrared structure
(long--distance behaviour) of the theory.
Rough fluctuations can also exhibit themselves in the form of
$~Z_{N}-$ strings and $~Z_{N}-$ monopoles [\ref{z2a}-\ref{z2b}]
($~Z_{N}-$artifacts).
In the paper \cite{bcm} the influence of small--size $~Z_{2}$-artifacts
(negative plaquettes) on `physical' values was studied
for pure gauge $~SU(2)~$ theory.
It was shown that these $~Z_{2}-$artifacts can strongly influence
large--scale objects, e.g., Wilson loops and their ratios.
In this connection it could be instructive to note the following.
The number of artifacts (e.g., negative plaquettes)
$~N_{-}~$ decreases exponentially with
increasing $~\beta~$. But this in itself does not mean that they
become  unimportant at large $~\beta~$. They can still strongly
influence any lattice average $~ \langle O(U) \rangle (\beta) ~$
(say, Wilson loops) decaying faster then $~N_{-}(\beta)~$ with
increasing $~\beta~$, and, therefore, even a small admixture
of these rapidly varying fields can become  competitive.

The question of contamination from lattice artifacts
should be addressed also to the pure gauge $~U(1)~$ lattice
theory which is expected to be the free photon's theory in the
continuum limit.
During last years this theory was rather well studied
(for a noncomplete list of references see, for instance,
[\ref{u1a}-\ref{u1b}]).
On the lattice the pure gauge $~U(1)~$ theory in four
dimensions with Wilson action has two phases separated
by a first order phase transition. In the strong
coupling phase monopoles (i.e., artifacts) are supposed to
form the condensate which causes
the confinement of electric charges. At the phase transition
point $~\beta_{c}~$ this condensate dissolves, and the
transition to the Coulomb phase occurs.
This theory has a nontrivial spectrum which differs
within these phases.
Its $~0^{++}~$, $~1^{+-}~$ and $~2^{++}~$ states were
studied in \cite{bp}.

To recover the continuum limit of this theory
it is important to study the role of the underlying
singular configurations to be separated from the `real' physics.
One can modify the lattice action
in such a way that the unphysical short--range fluctuations
(monopoles, $\ldots$) become suppressed leaving the `physical'
fields untouched. The question of interest is then whether
this modified theory can serve as a better lattice
approximation to that of noninteracting photons in the continuum.
To study this question in detail is the goal of the present work.

The paper is organized as follows.
First we discuss the lattice monopoles and
the modification of the action.
Section 3 is devoted to the discussion of correlators and
`glueball' wave functions.
Section 4 presents the results of Monte Carlo calculations and
their analysis.
The last section contains a discussion of the results as well as
some speculations.

\vspace{0.2cm}

\section{Monopoles and modified lattice action}

The standard Wilson action (WA) for the pure gauge
$~U(1)~$ theory is

\eq
 S_{W}(U_{l}) =
\beta \cdot \sum_{x;\mu \nu}
        \,  \bigl( 1 - \cos \theta_{x;\, \mu \nu} \bigr) ~,
                                              \label{wa}
\en

\noi where $~\beta = 1/g^{2}_{bare} \equiv 1/g^2~$, and
$~U_l \equiv U_{x \mu} = \exp (i \theta_{x;\, \mu} ),
\quad \theta_{x;\, \mu} \in (-\pi, \pi] ~$
are the field variables defined on the links $l = (x,\mu)~$.
Plaquette angles $~\theta_{P} = \theta_{x;\, \mu \nu}~$
are given by

\eq
\theta_{x;\, \mu \nu} =
  \theta_{x;\, \mu} + \theta_{x + \hat{\mu};\, \nu}
- \theta_{x + \hat{\nu};\, \mu} - \theta_{x;\, \nu} ~.
\en

To extract magnetic monopoles the plaquette angle
$~\theta_{P}~$ is split up

\eq
\theta_{P} = \overline{\theta}_{P} + 2\pi n_{P},
\quad \mbox{}-\pi < \overline{\theta}_{P} < \pi,\;\;\;
\quad n_{P} = 0, \pm 1, \pm 2~,
                                              \label{thetabar}
\en

\noi where $\overline{\theta}_{P}$ describes the electromagnetic
flux through the plaquette and $n_{P}$ is the number of Dirac
strings passing through it.
The net number of Dirac strings going out of an elementary
$~3D~$ cube determines the monopole charge within this cube as follows.
The monopole current out of the cube $c_{n,\mu}$, labeled by the
dual link ($n,\mu$), is defined to be \cite{dgt}

\eq
K_{n,\mu} =
\frac {1}{4\pi} \varepsilon_{\mu \nu \rho \sigma}
    \bigtriangleup_{\nu} \overline{\theta}_{n,\rho \sigma} ~,
                                                   \label{kmu1}
\en

\noi where the lattice derivative $~\bigtriangleup_{\nu}~$ is
defined by $~\bigtriangleup_{\nu}f_n=f_{n+\hat{\nu}}-f_n~$.
Then we have

\eq
 2\pi \, K_{n,\mu}= \sum_{P} \overline{\theta}_{P} =
 \sum_{P} (\theta_{P} - 2\pi n_{P})=
 -2 \pi \sum_{P} n_{P}~,
                                                    \label{kmu2}
\en

\noi where the sums are over oriented plaquettes $~P~$ in the
surface of the cube $~c_{n,\mu}~$.

\eq
\sum_{\mu} \bigtriangleup_{\mu} K_{n,\mu} = 0,       \label{210}
\en

\noi which means that the monopole currents form closed loops on
the dual lattice.

To suppress monopoles one can use the modified action (MA)

\eq
 S_{MA} = S_{W} + \lambda_{K} \cdot \sum_{c} \mid \! K_{c} \! \mid ~,
                                              \label{ma}
\en

\noi where the parameter $~\lambda_{K}(\beta)~$ plays the
role of a chemical potential for monopoles. In our calculations
we have chosen $~\lambda_{K} = \infty ~$. The partition function is

\eq
  Z_{MA} = \int  \! \! \prod_{links}  \! \! dU_{\em l}
{}~\prod_{c} \delta_{K_{c};0} \cdot \exp[~-S_{W}(U_{\em l})~]
                                                    \label{pf}
\en

\noi with periodic boundary conditions, and the average of
any field functional $~O(U)~$ is defined as

 \eq
 \langle O(U) \rangle_{MA}  = Z^{-1}_{MA}
 \cdot  \! \int \! \! \prod_{links}
 \! \! dU_{\em l}~O(U) \cdot \prod_{c} \delta_{K_{c};0} \cdot
 \exp[~-S_{W}(U_{\em l})~]~.
\en

\noi The action $~S_{MA}~$ was used in \cite{bss,bs} in
order to investigate the behaviour of the average
plaquette $~\langle P \rangle ~$.

For some $~\beta$--values we employed a further modification
of the action (MA1) suppressing all lattice artifacts
characterized by negative plaquette values

\eq
 S_{MA1} = S_{MA} +
 \lambda_{P} \cdot \sum_{P}
\Bigl( 1 - \sign (\cos \theta_{P}) \Bigr)~,
 \quad \lambda_{P} = \infty ~.
                                              \label{ma1}
\en

\noi Note that both modifications (MA and MA1) do not
influence the formal continuum limit and do not change any
perturbative aspects.

\vspace{0.5 cm}
\section{Correlators and boson wave functions}

Energies of the excited states (`glueballs') can be calculated
by measuring the asymptotic behaviour of the correlation
functions of operators $~\Phi~$ with the proper
$J^{PC}$ quantum numbers and momenta $~\vec{p}~$.

The corresponding correlation functions $~\Gamma (\tau)~$
on a $~L_{t} \cdot L_{s}^{3}~$ lattice are defined as follows

\eqa
\Gamma (\tau)
 & = & \langle \, \Phi^{\ast}(t+\tau) \cdot \Phi(t) \, \rangle^{c}
\nonumber \\
 & \equiv & \nn \sum_{i=1}^{n} \, \frac{1}{L_{t}} \sum_{t=0}^{L_{t}-1}
\left[ \Phi_{i}^{\, \ast}(t \oplus \tau) \cdot \Phi_{i}(t)
- \overline{\Phi}^{\, \ast}(t \oplus \tau) \cdot
\overline{\Phi}(t) \right] ~,
                                         \label{gamma1}
\ena

\noi where $~n~$ is the number of measurements,
$~t \oplus \tau = (t+\tau) \, \mbox{mod} (L_{t})~$ and

\eq
\overline{\Phi}(t) = \n \sum_{i=1}^{n} \, \Phi_{i}(t) ~.
                                         \label{phibar}
\en

\noi This definition provides us with unbiased estimators for
$~\Gamma (\tau)~$.

The correlators defined above are the superpositions of
the exponents corresponding to different energy levels, e.g.,

\eq
 \frac{ \Gamma (\tau)}{ \Gamma (0)} =
  A \cdot \left[ e^{-E \cdot \tau} +
  e^{-E \cdot (N_{t}-\tau)} \right] + \ldots ~,
                                                  \label{gamma2}
\en

\noi where dots correspond to higher state contributions
and

\eq
A = \frac{1}{ \Gamma (0)} \cdot
\mid \! \langle J^{PC}  \! \mid  \Phi^{J^{PC}}(0) \mid \! 0 \rangle \!
\mid^{\, 2} ~\leq 1 ~.
                                                  \label{a1}
\en

In principle the lowest state energy can be defined from
the ratio of correlators at large enough $~\tau~$
(but $~\tau \ll L_{t}/2~$) :

\eq
 E_{eff} \equiv - \ln \frac {\Gamma(\tau)}{\Gamma(\tau-1)}~,
                                                  \label{en1}
\en

\noi and  $~E = E_{eff} \quad
\mbox{if} \quad E \! \cdot \! (\tau -1) \gg 1~$.
In practice only for comparatively small values of energy
($E \sim 1-2$) we can use the separation distance  up to $~\tau = 5~$.
For larger values of energy the  correlators decay very
quickly, the signal--to--noise ratio becomes too bad,
and at $~E \ageq 4~$ it is possible to use only $~\tau = 0~$
and $~\tau = 1~$ values with reasonable statistical errors.
In this case the effective energy $~E_{eff}(\tau)~$
in eq.(\ref{en1}) is nothing but an estimate of the
 true energy $~E~$. The question how close $~E_{eff}(\tau)~$ is
to the true value $~E~$ is the question of contamination from
higher energy states. The larger the overlap of our `glueball'
wave function with the lowest lying `glueball' state, the
closer $~A~$ is to $~1~$, the less the contamination from
higher energy states.
The common way to suppress the higher states contribution
is to consider 'smeared' wave functions which
are linear combinations of large numbers of Wilson loops and
have a large overlap with low lying states \cite{tep,ape,bks}.

In our work we used the simplest plaquette wave functions
which are known to have comparatively small overlaps with
the lowest glueball states (at least for the standard Wilson action).
For us this is not the point of great
importance because our main goal is the comparison of predictions
for the different actions. Also the plaquette wave
functions are proved to work very well in the case of comparatively
small energies (where we can use the larger values of $~\tau$).
We shall come back to this question lateron, when discussing
the influence of $~Z_{2}$--artifacts.

To construct the one-plaquette operators we followed the
standard recipe described elsewhere (for a review see,
for instance, \cite{ist}).
The first step is to construct local operators $~\Phi(t;\vec{r} \, )~$
whose $~\vec{p}=0-$parts have the desired $~J^{PC}~$ properties.
Real parts of Wilson loop operators have $~C-$parity
$~C=+1$ and imaginary parts have  $~C-$parity  $~C=-1$.
For the $~0^{++}~$ state we first sum up the real parts of
all six plaquettes belonging to the smallest space--like cube

\eq
\Phi(t;\vec{r} \, ) = \frac{1}{6} \sum_{k=1}^{6}
\cos (\theta_{P_{k}}) ~,
\en

\noi where  $~\vec{r}~$ is the centre of the cube.
This construction is invariant under  $~\pi/2~$ and
 $~\pi~$  rotations about all three axes in the $~3D~$ space
passing through its centre. It is easy to show that
the zero momentum operator

\eq
\Phi(t) = \sum_{\vec{r}} \Phi(t;\vec{r} \, )
                                   \label{phi1}
\en

\noi corresponds to $~J^{PC}=0^{++},~4^{++},~\ldots~$ exitations.
We assume that the state with $~J=0~$ has the minimal mass,
so, we shall refer to it as the $~0^{++}~$ state.

The Fourier transform with momentum $~\vec{p}~$ gives us
non-zero momentum operators:

\eq
\Phi(t;\vec{p}) = \sum_{\vec{r}}
e^{-i\vec{r}\vec{p}} \cdot \Phi(t;\vec{r})
                                   \label{ft}
\en

For constructing a tensor operator one has to take plaquettes parallel
to the quantization axis (say, $~OY$) and subtract from it
the same but rotated by $~\pi/2~$ about $~OY~$. After summing
through the three--dimensional space we obtain the operator
corresponding to $~J^{PC}=2^{++},~4^{++},~\ldots~$ excitations.
As far as it is commonly assumed that the state with $~J=2~$
has the minimal mass it is refered as the $~2^{++}~$ state
with spin projection $~\pm 2~$ onto the quantization axis.
In this case the momentum $~\vec{p}~$ in eq.(\ref{ft}) must
be chosen parallel to the quantization axis to avoid the
admixture of a scalar state.

Finally, summing up the imaginary parts of plaquette
operators  parallel to the quantization axis results
in $~1^{+-},~3^{+-},~\ldots~$ states (or just $~1^{+-}$ state).
Comparing with the photon it has the wrong parity at zero momentum.
But at $~\vec{p} \neq 0~$ it has a non-zero overlap with
the $~1^{--}~$ (photon) state \cite{bp}. This fact can be easily
checked in the naive continuum limit.

\vspace{0.5cm}

\section{Monte Carlo calculations}

 Our calculations were made on a
$~L_{t} \cdot L_{s}^{3} = 12 \cdot 6^{3}~$ lattice at
$~0.1 \leq \beta \leq 2~$.
For each $~\beta~$ value our statistics reaches from $~40.000~$
to $~50.000~$ sweeps, and $~2000~$ sweeps without measurements were
done to reach equilibrium.
For calculating the mean square errors and biases of our estimators
we used the jackknife method \cite{efron,yr}.

We calculated autocorrelation functions $~C^{auto.}(n_{sweep})~$
for all our observables (plaquettes $~P_{s,t}~$,
correlators $~\Gamma (\tau;\vec{p})~$, etc.) and
defined the decorrelation length $~n_{dec.}~$ as a
minimal number of sweeps such that

\eq
 C^{auto.}(n_{dec.}) < \frac{1}{5} \cdot C^{auto.}(0) ~.
\en

\noi For the standard Wilson action $~n_{dec.}~$
gets a sharp peak at $~\beta = \beta_{c} \sim 1~$
with $~n_{dec.} \ageq 100~$, while for
the modified actions this value is comparatively
small ($n_{dec.} \sim 4-6$) and has a very weak
dependence on $~\beta~$ in the chosen interval.

In Fig.1 we show the dependence of the average plaquette
$~\langle P \rangle ~$ on  $~\beta~$ for WA
and MA. For the modified action
this dependence is very smooth
and shows no trace of crossover or phase transition in agreement
with the calculations of \cite{bss,bs}.

Another important parameter which we care about
is the average number of negative plaquettes
$~\langle N_{-} \rangle ~$, i.e., plaquettes $~P_{x;\mu \nu}~$
with

\eq
 \cos (\theta_{x;\, \mu \nu}) < 0~.
\en

\noi In Fig.2 we show the dependence of its density
$~\rho_{neg.} \equiv \langle N_{-} \rangle /N_{P}~$ on
$~\beta~$ for WA and MA.
It is interesting to note that the suppression of monopoles
decreases the density of negative plaquettes
at $~\beta \aleq 1.1~$, while at larger $~\beta~$ values
$~\rho_{neg.}~$ is approximately the same for both the actions.

As an example in Fig.3 we show the  $~\tau~$ dependence
 of the vector state correlator $~\Gamma(\tau)~$
at nonzero momentum $~\vec{p} = (0,1,0) \cdot \pi/3a~$
at $~\beta=0.3~$ and $~\beta=1.0~$ for WA and MA. For the
modified action the correlator does not depend on the coupling
giving an indication for the decoupling of the photon. Because
the energy is comparatively small these correlators have
reasonable errors up to $~\tau = 5~$. On the contrary, for
the Wilson action at $~\beta = 0.3~$ we recover the
expected strong fall--off of the correlator.

In order to extract masses from the effective energy
$~E_{eff}(\vec{p};\tau)~$ we need a dispersion relation (DR).
In the free field boson case the propagator on the lattice
provides us with the DR

\eq
 \sinh^{2} \frac{a E}{2} = \sinh^{2} \frac{a m}{2} +
\sum_{i=1}^{3} \sin^{2} \frac{a p_{i}}{2} ~,
                                                \label{ldr}
\en

\noi which at small enough spacing and momenta gives us the continuum DR:

\eq
 E^{2} = m^{2} + \vec{p}^{\, 2}~.
                                                \label{cdr}
\en

\noi In general the connection between the momentum and
the energy can be very nontrivial on lattice,
and a DR which we can expect is of the form (see, e.g., \cite{ku})

\eqa
 \sinh^{2} \frac{a E}{2}
& = & z_{1}(\beta;J^{PC}) \cdot \sinh^{2} \frac{a m}{2} +
z_{2}(\beta;J^{PC}) \cdot
\sum_{i=1}^{3} \sin^{2} \frac{a p_{i}}{2}
\nonumber \\
& & + z_{3}(\beta;J^{PC}) \cdot
\sum_{i,j=1}^{3} \frac{1}{a^{2}} \cdot
\sin^{2} \frac{a p_{i}}{2} \cdot \sin^{2} \frac{a p_{j}}{2}
+ \ldots ~.
                                                \label{ldr1}
\ena

\noi We believe that in the continuum limit

\eqa
 z_{i}(\beta;J^{PC}) & \ra & 1 \quad \mbox{at} \quad i =1,2~;
\nonumber \\
 z_{i}(\beta;J^{PC}) & \ra & 0 \quad \mbox{at} \quad i \geq 3~.
                                                     \label{zi}
\ena

\noi The second line in eq.(\ref{zi}) provides the necessary
condition for the restoration of rotational invariance
(at small enough $~\vec{p}~$), and the
first one yields the necessary condition for the restoration
of Lorentz invariance.

In Figs.4 we show the effective energy of the vector boson
$~E_{eff}(\vec{p};\tau)~$ as a function of $~\beta~$
for Wilson action (Fig.4a) and for the modified action (Fig.4b).
Different symbols correspond to different momenta $~\vec{p}~$
as indicated in the figures.
Broken lines correspond to the free field theory dispersion
relation eq.(\ref{ldr}) with zero mass and different momenta
$~\vec{p}~$.
In the theory with the MA a massless photon exists
in the whole $~\beta~$ region we have chosen, and the free
field theory dispersion relation (\ref{ldr}) works
astonishingly well.
Therefore, it provides a very convincing evidence that
in the theory without monopoles a massless photon exists even at very
small $~\beta~$.
In the theory with the WA a massless photon exists in the Coulomb phase
as already observed in \cite{bp} (see also \cite{cc}) while
below the critical point the photon acquires a nonzero mass.

Another interesting observation is that the DR (\ref{ldr})
works very well for both the actions at $~\beta \ageq 1~$
even for $~\tau = 1~$. This fact shows that the overlap
coefficient $~A(\beta)~$ is close enough to unity
in this phase, and the contamination from higher states becomes
negligible with increasing $~\beta~$.

Analogously to Figs.4 in Figs.5 we show the effective energy
$~E_{eff}(\vec{p};\tau =1)~$ of the $~0^{++}~$ boson
as a function of $~\beta~$
for the Wilson action (Fig.5a) and modified action (Fig.5b).
For the MA the dependence on the coupling is very smooth, and
at increasing $~\beta~$ the energies for both actions tend to be
rather close to each other.
It is difficult to check the validity of the free field
theory DR with our data because the energies are comparatively big,
and their dependence on the momentum is weak.

Similar observations can be made for the tensor states.
In Figs.6 we show the effective energy $~E_{eff}(\vec{p};\tau =1)~$
of the $~2^{++}~$ boson as a function of $~\beta~$
for the Wilson action (Fig.6a) and the modified action (Fig.6b).

Masses $~m^{\gamma}~$  we calculated are in units
of the (unknown) spacing $~a(\beta)~$:

\eq
  m^{\gamma} = m^{\gamma}_{ph.} \cdot a(\beta)~;
\qquad \gamma=0^{++}, \ldots~;
\en

\noi but their ratio

\eq
  m^{2^{++}}/m^{0^{++}} = m^{2^{++}}_{ph.}/m^{0^{++}}_{ph.}
\en

\noi must not depend on the coupling in the continuum limit.
Comparing the data for scalar and tensor states it is easy to see
that this ratio has a very weak $~\beta~$ dependence in the whole
interval and is rather close to unity.

As it was shown above the simple one-plaquette wave functions
work very well for $~\tau \geq 2~$ (at least, for the vector state).
This means that the overlap coefficient $~A(\beta)~$ which can be
defined as

\eq
   A(\beta) = \frac{\Gamma^{2}(\tau =1)}{\Gamma(\tau =2)} ~.
\en
\noi is not too far from unity.
Fig.7a shows its dependence on $~\beta~$ for the vector state at
$~\vec{p} = (0;1;0) \cdot \pi/3a~$.
For MA (open circles) $~A \sim 0.65~$ at small $~\beta~$,
and at $~\beta \sim 1~$ it becomes close to $~1~$.
It is tempting to assume that there is some correlation between
this dependence and the $~\beta-$dependence of
the number of negative plaquettes
(compare with Fig.2).
To check this assumption we used another modified action (MA1)
with the full suppression of negative plaquettes (\ref{ma1}).
Calculations were made at $~\beta =0.3~$ and $~\beta =0.5~$, and
the corresponding values of $~A(\beta)~$ are shown in Fig.7a
with full circles.
In Fig.7b we show the effective energy
$~E_{eff}^{J=1}(\vec{p};\tau =1)~$ at $~\beta =0.3~$
as a function of $~\hat{p}^{\, 2} = \sum_{i=1}^{3} 4 \,
\sin^{2} \left( a p_{i} /2 \right)~$.
Open circles correspond to MA, and
full circles correspond to MA1.
The solid line corresponds to the dispersion relation (\ref{ldr}).
Therefore, after suppressing negative plaquettes the agreement
between the $~\tau =1~$ effective energy and the dispersion relation
(\ref{ldr}) becomes practically perfect. So, we conclude
that the $~Z_{2}$--artifacts are, at least partly,
responsible for not good enough overlap with the higher states.
It is worthwhile to note here that the suppression of
negative plaquettes does not influence the energy of the photon
while the energies of the massive scalar, vector and tensor states
tend to increase.

\section{Conclusions}

We have studied the properties of the compact
four--dimensional $U(1)$ lattice gauge theory without monopoles.
Employing Monte Carlo simulations we calculated correlators
of scalar, vector and tensor operators at zero and nonzero
momenta $~\vec{p}~$ on a $~12 \cdot 6^{3}~$ lattice
at $~0.1 \leq \beta \leq 2~$.
We confirm that the theory without monopoles has no phase
transition, at least, in the interval of $~\beta~$ studied.
In the modified theory a massless photon exists in the whole interval
of $~\beta~$, fitting the lattice free field theory
dispersion relation very well.
The energies of  $~0^{++}~$, $~1^{+-}~$ and $~2^{++}~$ states
show a rather weak dependence on the coupling in the chosen interval
of $~\beta~$, and their ratios are practically constant.
We show also that other lattice artifacts
-- negative plaquettes -- are, at least partly,
responsible for the overlap in the strong coupling region.
The further modification of the theory by suppressing
the negative plaquettes drastically improves the overlap
with the lowest states (at least, for $~J=1$).

We believe that the problem of artifacts in lattice calculations
deserves much more detailed and thorough study,
and still our understanding of pure gauge $~U(1)~$ theory is far from
being complete.
Even after suppressing monopoles (and negative plaquettes)
there is still some residual interaction which occurs in
the appearence of massive states ($0^{++}, \ldots $).
One can speculate that there are other kinds of lattice artifacts
which are responsible for this interaction, and Dirac strings are
among the possible candidates for a further study.

We conclude that the modified theory without lattice artifacts
(monopoles, negative plaquettes, etc.) should be a more
reliable candidate for the QED calculations with fermions
in the strong coupling area.

\vspace{0.5 cm}

\begin{center}
\noindent {\large \bf Acknowledgements.}
\end{center}

One of us (V.G.B.) would like to express his deep
gratitude to all collegues at Humboldt University for hospitality.

\newpage
\begin{center}
\noindent {\large \bf Figure captions}
\end{center}

\vspace{.5cm}

\noi {\bf Fig.1}  The plaquette value $~\langle P \rangle ~$
for WA (squares) and MA (circles) as a function of $~\beta~$.

\vspace{0.5cm}

\noi {\bf Fig.2} The density of negative plaquettes
$~\rho_{neg.}~$ for WA (squares) and MA (circles) vs. $~\beta~$.

\vspace{0.5cm}
\noi {\bf Fig.3} Correlator $~\Gamma(\tau;\vec{p})~$ for the
vector state at $~\beta=0.3~$ and
$~\beta=1.0~$ for WA and MA.
$~\vec{p} \equiv \vec{k} \cdot 2\pi/aL_{s}~$

\vspace{0.5cm}

\noi {\bf Fig.4} The effective energy of the vector state
$~E_{eff}(\beta;\tau)~$ at different
$~\vec{p} \equiv \vec{k} \cdot 2\pi/aL_{s}~$ as a function of $~\beta~$
for WA ({\bf a}) and for MA ({\bf b}).
Different symbols correspond to different momenta
$~\vec{p}~$ shown in the picture.
Broken lines correspond to the dispersion relation (\ref{ldr})
with zero mass.

\vspace{0.5cm}

\noi {\bf Fig.5} The effective energies $~E_{eff}(\vec{p};\tau =1)~$
for the scalar state as a function of $~\beta~$
for WA ({\bf a}) and for MA ({\bf b}).

\vspace{0.5cm}

\noi {\bf Fig.6} The effective energies $~E_{eff}(\vec{p};\tau =1)~$
of the tensor state as a function of $~\beta~$
for WA ({\bf a}) and for MA ({\bf b}).

\vspace{0.5cm}

\noi {\bf Fig.7} $~A(\beta) = \Gamma^{2}(\tau =1) / \Gamma(\tau =2)~$
vs. $~\beta~$ for the vector state for MA and MA1 ({\bf a});
The effective energy $~E_{eff}(\vec{p};\tau =1)~$ of the
vector state at $~\beta =0.3~$
as a function of $~\hat{p}^{2} = \sum_{i=1}^{3} 4 \,
\sin^{2} \left( a p_{i} /2 \right)~$ for MA and MA1.
The momenta are the same as in Fig.4.
The solid line corresponds to the dispersion relation (\ref{ldr})
({\bf b}).


\newpage
\begin{thebibliography}{99}
\newcommand{\prd}[1]{Phys.~Rev.~{\bf D#1}\ }
\newcommand{\plb}[1]{Phys.~Lett.~{\bf #1B}\ }
\newcommand{\npb}[1]{Nucl.~Phys.~{\bf B#1}\ }
\newcommand{\prl}[1]{Phys.~Rev.~Lett.~{\bf #1}\ }
\newcommand{\pr}[1]{Phys.~Rep.~{\bf #1}\ }
\newcommand{\ap}[1]{Ann.~Phys.~{\bf #1}\ }
\newcommand{\cmp}[1]{Commun.~Math.~Phys.~{\bf #1}}
\newcommand{\rmp}[1]{Rev.~Mod.~Phys.~{\bf #1}}
\newcommand{\ptp}[1]{Prog.~Theor.~Phys.~{\bf #1}}
%
\bibitem{qed1}  G.~Schierholz,
              ~~Nucl. Phys. {\bf B} (Proc. Suppl.) {\bf 20} (1991) 623
\bibitem{qed2}  E.~Dagotto, in {\it Dynamical Symmetry Breaking},
                ed. K.~Yamawaki, World Scientific (1992) p.189
\bibitem{wang}  K.~C.~Wang,
              ~~Nucl. Phys. {\bf B} (Proc. Suppl.) {\bf 20} (1991) 646
\bibitem{pt1} D.J.R.~Pugh and M.~Teper,  ~~Phys. Lett. {\bf B218} (1989) 326
\bibitem{pt2} D.J.R.~Pugh and M.~Teper,  ~~Phys. Lett. {\bf B224} (1989) 159
\bibitem{desy1} M.~G\"{o}ckeler, A.S.~Kronfeld, M.L.~Laursen,
              G.~Schierholz and U.-J.~Wiese,
              ~~Phys. Lett. {\bf B233} (1989) 192
\bibitem{lus2} M.~L\"{u}scher, ~~Nucl. Phys.  {\bf B200} (1982) 61
\bibitem{hkp} K. Huang, J. Koike and J. Polonyi,
             ~~Int. J. Mod. Phys. {\bf A6} (1991) 1267
\bibitem{hp}  K. Huang and J. Polonyi,
             ~~Int. J. Mod. Phys. {\bf A6} (1991) 409
\bibitem{mp} G.~Mack and V.~Petkova, ~~Ann. Phys. {\bf 123} (1979) 442;
             ~~Ann. Phys. {\bf 125} (1980) 117               \label{z2a}
\bibitem{mpiet} G.~Mack and E.~Pietarinen, ~~Nucl. Phys.
              {\bf B205}  (1982) 141
\bibitem{yaf} L.~Yaffe, ~~Phys. Rev. {\bf D21} (1980) 1574
\bibitem{tom} T.~Tombolis, ~~Phys. Rev. {\bf D23} (1981) 2371
\bibitem{hs} I.~Halliday and A.~Schwimmer, ~~Phys. Lett. {\bf B102} 1981 337
\bibitem{bkl1} R.C.~Brower, D.A.~Kessler and H.~Levine, ~~Nucl. Phys.
              {\bf B205}  (1982) 77                          \label{z2b}
\bibitem{bcm} V.G.~Bornyakov, M.~Creutz and  V.K.~Mitrjushkin,
                  Phys. Rev. {\bf D44} (1991) 3918
\bibitem{kss} J. Kogut, D. R. Sinclair and L. Susskind,
              \npb{114} (1976) 199                           \label{u1a}
\bibitem{bmk} T. Banks, R. Myerson and J. Kogut,
              \npb{129} (1977) 493
\bibitem{dgt}  T.A.~DeGrand and D.~Toussaint,
               ~~\prd{22} (1980) 2478
\bibitem{bp}  B. Berg and C. Panagiotakopoulos, ~~Phys. Rev. Lett.
             {\bf 52} (1984) 94
\bibitem{ejnz} H. G. ~Evertz, T.~Jers\'{a}k, T.~Neuhaus and
               P. M.~Zerwas,  \npb{251} (1985) 279
\bibitem{bss} J. S. Barber, R. E. Schrock and R. Schrader,
             ~~~Phys. Lett. {\bf B152} (1985) 221
\bibitem{bs}  J. S. Barber and R. E. Schrock,
              \npb{257} (1985) 515
\bibitem{wens}  S.~Hands and  R.~Wensley, ~~Phys. Rev. Lett.
                {\bf 63} (1989) 2169
\bibitem{pw}  L.~Polley and U.-J.~Wiese,
              \npb{356} (1991) 629
\bibitem{sw}  J.~D.~Stack and R. J. Wensley,
              \npb{371} (1992) 597
\bibitem{kl}  H.~Kleinert,
              Freie Univ. Berlin preprint,  July 1992
\bibitem{cc} P. Cea and L. Cosmai,
               Bari-preprint BARI-TH 89/91 (1991)            \label{u1b}
\bibitem{ist} K. Ishikawa, G. Schierholz and M. Teper,
              Z. Phys {\bf C19} (1983) 327
\bibitem{tep} M. Teper,
             ~~~Phys. Lett. {\bf B183} (1987) 345
\bibitem{ape} The APE collaboration, M. Albanese et al.,
             ~~~Phys. Lett. {\bf B192} (1987) 163
\bibitem{bks} F.~Brandstaeter, A.~S.~Kronfeld and G. Schierholz,
               Nucl. Phys {\bf B345} (1990) 709
\bibitem{ku}  N.~Kimura and A.~Ukawa,
              \npb{205} (1982) 637
\bibitem{bb}  B.~A.~Berg and  A.~H.~Billoire, ~~Phys. Rev. {\bf D40}
              (1989) 550
\bibitem{efron} R.~G.~Miller, Biometrika {\bf 61} (1974) 1;
               B. Efron, SIAM Review {\bf 21} (1979) 460
\bibitem{yr}   M. C. K. Yang and D. H. Robinson, `Understanding and
              learning statistics by computer', World Scientific, serie in
              computer science, Vol. {\bf 4} (1986)
\end{thebibliography}
\end{document}